\documentclass[aps,pra,twocolumn,showpacs]{revtex4}
\begin{document}
\title{One-dimensional optical lattices and impenetrable bosons}
\author{M. A. Cazalilla}
\affiliation{The Abdus Salam ICTP,
Strada Costiera 11, 34014 Trieste, Italy,}
\affiliation{ Donostia International Physics Center (DIPC),
Manuel de Lardizabal 4, 20018 Donostia, Spain.}
\begin{abstract}
We study the limit of large onsite repulsion of the one-dimensional
Bose-Hubbard model  at low densities, and derive a strong-coupling
effective Hamiltonian. By taking the lattice parameter to zero,  the 
Hamiltonian  becomes 
a continuum model of fermions  with attractive interactions. The leading
corrections to the internal energy of a hard-core-boson (Tonks) gas as
well as the (finite temperature) 
pair correlations of a strongly interacting Bose gas are calculated.
We  explore the possibility of realizing,  in an optical lattice, a
Luttinger liquid  with  stronger 
density correlations than the Tonks gas. A quantum phase transition to 
a charge-density-wave Mott insulator is also discussed.  
\end{abstract}
\maketitle
The field  of ultra-cold atoms is  attracting 
much attention in recent times. This is  partly due to the possibility
(in some cases already experimentally demonstrated) of realizing
exotic quantum phase transitions~\cite{Sachdev00}.  One of the most
interesting examples is  the transition between  a superfluid and a Mott 
insulator recently observed 
in an optical lattice~\cite{Greiner02}.  The prototypical model for this
transition is the Bose-Hubbard model, which in one-dimension takes the 
form:
\begin{equation}\label{bhm}
H_{BH} = -\frac{J}{2} \sum^{M_{\rm o}}_{m=1} \left[b^{\dag}_{m+1}b_{m} +
{\rm h.c.} \right] 
+ \frac{U}{2} \sum^{M_{\rm o}}_{m=1} n_m (n_m-1),
\end{equation}
where $[ b_{m}, b^{\dag}_{m'}] = \delta_{m,m'}$, and commute otherwise;
$n_{m} = b^{\dag}_{m} b_{m}$ counts the number of bosons at the $m$-th
site, where
$m = 1,\ldots, M_{\rm o}$  ($M_{\rm o}$ is the number of lattice sites).

 As it has been  recently discussed by B\"uchler {\it et
al.}~\cite{Buchler02}  (see also~\cite{Haldane81,Fisher89}),  in one-dimension 
the superfluid-Mott  insulator transition can be described by  the  sine-Gordon model:
\begin{equation}\label{sgm}
H_{\rm SG} =  H_{\rm gauss}[\theta,\phi,K] + g \int^{L}_{0} dx\:  \cos
\left[ 2\theta(x)  +2 \pi x \delta  \right].
\end{equation}
where
\begin{equation}\label{hgauss}
H_{\rm gauss} = \frac{\hbar v_s}{2\pi}\int dx\:
\left[\frac{\left(\partial_x \theta  
\right)^2}{K} + K \left(\partial_x \phi \right)^2\right]. 
\end{equation}
describes the (long wave-length) gaussian 
fluctuations of  the phase $\phi(x)$ and density $\theta(x)$
fields~\cite{Efetov76,Haldane81};
$\delta = \rho_{\rm o} - a^{-1}$ is the mismatch
between the mean density, $\rho_{\rm o}$, and the inverse lattice
parameter $a^{-1}$.
The sine-Gordon model exhibits  a Kosterlitz-Thouless (KT)
transition~\cite{Fisher89,Buchler02}
close to $K = 2$ for $\delta = 0$, and a commensurate-incommensurate 
transition  for $K \leq 2$  as $\delta$ is varied~\cite{Nersesyan}.  The
dimensionless
parameter $K$ is proportional to the compressibility~\cite{Cazalilla02}, 
and controls the spatial decay of
phase and density correlations. 
Since in one dimension Bose-condensation cannot take
place even at zero temperature, correlations decay at least
algebraically. Thus,  for a  sufficiently large system at $T = 0$,  the
one-body density matrix displays
a power-law behavior~\cite{Efetov76,Haldane81}: 
$\langle \Psi^{\dag}(x) \Psi(0) \rangle \sim |x|^{-1/2K}$. Density
correlations, however, decay 
as~\cite{Efetov76,Haldane81} $\langle \rho(x) \rho(0) \rangle \sim x^{-2} 
+ A \: \cos(2 \pi\rho_{\rm o}) |x|^{-2K} $, where  $A$ is a  constant. For
a zero-range interaction, 
$v(x) = g\,  \delta(x)$, Lieb~\cite{Lieb65} found the exact eigenstates
and eigenvalues of the Hamiltonian
using the Bethe ansatz. This allows  to obtain $K$  for every value of 
the dimensionless parameter $\gamma  =   M g/ \hbar^2 \rho_{\rm o}$ ($M$
being
the boson mass).  However, for this model 
$K \geq 1$ always, and takes the smallest possible
value ($K = 1$) as $\gamma \to +\infty$ and  the  system  becomes a
Tonks gas~\cite{Girardeau65,Cazalilla02}. In this limit, many 
of the long-distance  correlations are the same as for a non-interacting 
Fermi  system~\cite{Efetov76,Haldane81,Cazalilla02}. 

In this paper we will consider the large $U$ limit of the Bose-Hubbard
model, Eq.~(\ref{bhm}), when the number of particles, $N_{\rm o}$, is less than the number of
lattice sites. 
We  show the following:
\begin{enumerate}
 \item
For $U \gg J$,   at  low temperatures, Eq.~(\ref{bhm}) can be effectively
replaced  
by  a model of {\it spinless} fermions  with correlated hopping and
attractive interactions:
\begin{eqnarray}\label{hfm}
&&H_F= -\frac{J}{2} \sum^{M_{\rm o}}_{m=1} \left[c^{\dag}_{m+1} c_{m} 
- \frac{J n_m}{U} c^{\dag}_{m+1}c_{m-1}  + {\rm h.c.} \right]
\nonumber \\ 
&& \mbox{     }- \frac{J^2}{2U} \sum_{m=1}^{M_{\rm o}} 
\Big[ n_{m+1} + n_{m-1} \Big] n_m + O\left(\frac{J^3}{U^2}\right),
\end{eqnarray}
where $\{ c_{m}, c^{\dag}_{m'}  \} = \delta_{m,m'}$, anti-commuting
otherwise, and 
$n_m = c^{\dag}_m c_m$.
\item
 When the lattice parameter $a \to 0$ and the Bose-Hubbard 
model becomes a continuum model of 
bosons interacting with zero-range interactions (the so-called
Lieb-Liniger model),  $H_F$ becomes~\footnote{After this work appeared 
in the archives, Dr. D. Sen  informed me that he has independently 
obtained a result equivalent to my  Eq.~(\ref{hfmc}), 
starting from the continuum (Lieb-Liniger)  model. See 
D. Sen, cond-mat/0301210 and Int. J. of Mod. Phys. A. {\bf 14}, 1789 (1999).}:
\begin{eqnarray}\label{hfmc}
&H_F&= \frac{\hbar^2}{2M} \int^{L}_{0}dx\:  \left(\partial_x \Psi_F^{\dag}
\right) 
\left(\partial_x \Psi_F \right) \nonumber \\
&-& \frac{g_F}{2} \int^{L}_{0} dx\: \partial_x \Psi^{\dag}_{F}(x)
\partial_x\Psi_F(x) 
\Psi^{\dag}_F(x) \Psi_F(x),
\end{eqnarray}
with the identifications  $\Psi_F(x_m = ma) = a^{-1/2} c_m$, $L= M_{\rm
o}a$, 
$g_F = 4\hbar^4/M^2g$, and $\hbar^2/M = J a^2$,  $g = Ua$. Using this
effective {\it fermionic} Hamiltonian we provide elementary derivations
for the $O(\gamma^{-1})$ correction 
to internal energy as well as for the (temperature-dependent) pair
correlation function:
$g_2 = \rho^{-2}_{\rm o}\: \langle \Psi^{\dag}(x) \Psi^{\dag}(x) \Psi(x)
\Psi(x) \rangle$, where
$\Psi(x)$ is the boson field operator.
\item
By including the interaction with the nearest neighbors, we argue that it
is possible 
to realize in an optical lattice a Luttinger liquid with $K < 1$. In this
regime a boson system resembles a 
system of fermions  with {\it repulsive} interactions. In particular, 
the system can undergo a quantum phase transition to 
a charge-density-wave Mott insulator in which 
the boson density is modulated with twice the period of the optical
lattice. 
\end{enumerate}

 We now proceed to derive the results listed above. Let us consider a
one-dimensional 
optical lattice. As usual, this presumes the existence of a tight
confinement 
in the transverse direction;  the effect of the confinement in the
longitudinal direction will be 
discussed at the end.  Assuming that $U$ is the largest
energy scale in the problem, i.e. $U \gg \max\{J, T\}$,   only states with  
site occupancy  equal or less than one (i.e. $n_m \leq 1$) will be
available because 
they have zero interaction energy. Thus we can split the Hilbert space
into two 
complementary  subspaces: ${\cal H} = {\cal H}_{P} \oplus {\cal H}_{Q} $.
We denote by $P$
the projector onto the subspace ${\cal H}_{P}$ of states 
which obey the constraint $n_m \leq 1$ for all $m$; 
$Q =  1 - P$  projects onto the complementary 
subspace, ${\cal H}_{Q}$. Our goal in what follows
is to find the effective Hamiltonian within the subspace ${\cal H}_P$. 
To this purpose we use the perturbative 
expansion (see e.g. ~\cite{Cohen92}):
\begin{equation}
H_{\rm eff} = PH_{BH} P - PH_{BH}Q \frac{1}{QH_{BH}Q} Q H_{BH} P + \cdots
\end{equation}
Here we shall not need the terms denoted by ``$\cdots$''  To $O(J^2/U)$ we
obtain:
\begin{eqnarray}\label{p1}
&&H_{\rm eff} \approx  -\frac{J}{2} \sum_{m=1}^{M_{\rm o}}
P(b^{\dag}_{m+1} b_{m} 
+ b^{\dag}_{m} b_{m+1} )  P \nonumber\\
&&- \frac{J^2}{4U}\sum_{m=1}^{M_{\rm o}} P(b^{\dag}_{m+1} +
b^{\dag}_{m-1}) b_{m} Q 
b^{\dag}_{m} (b_{m+1} + b_{m-1}) P.\nonumber\\
\end{eqnarray}
The first term in the previous expression describes the hopping of
impenetrable
bosons. The projection operators can be  removed if we represent the 
boson operators by Pauli matrices: $(b^{\dag}_m + b_{m}) \to
\sigma^{x}_m$, $i(b_{m}-b^{\dag}_{m} ) \to \sigma^{y}_{m}$,  and $2n_{m} - 1 \to
\sigma^{z}_{m}$.
Alternatively, as we shall do henceforth, one can use 
Jordan-Wigner fermions (see e.g. ~\cite{Sachdev00})
$b_{m} = \exp\big[ i\pi \sum_{l<m} n_l \big] \: c_{m}$,
and $n_m = b^{\dag}_{m}b_{m} = c^{\dag}_{m}c_{m}$, after restricting
ourselves to ${\cal H}_P$.  
In terms of the fermionic operators,
\begin{equation}\label{p2}
PHP = -\frac{J}{2} \sum_{m=1}^{M_{\rm o}} \left[ c^{\dag}_{m+1}c_m +
c^{\dag}_{m} c_{m+1}\right].
\end{equation}

 We now turn our attention to the second term in Eq.~(\ref{p1}). The
hopping terms
on the left side acting on $P$  move particles from the neighboring sites
to the $m$-th site. Therefore, in the resulting state, the $m$-th site can be either singly
our doubly occupied. The former possibility is projected out by the operator $Q$. Thus we can
replace $Q$  by $(n_m - 1)$, which does nothing when $n_m = 2$ and vanishes when $n_m =
1$. 
After some manipulations, we obtain:
\begin{eqnarray}
-\frac{1}{U}&(PH_{BH}Q)&(QH_{BH}P) = -\frac{J^2}{2U} \sum_{m=1}^{M_{\rm
o}} 
P  (b^{\dag}_{m+1}b_{m-1}  \nonumber\\ 
&&+ b^{\dag}_{m-1}b_{m+1} + n_{m+1} + n_{m-1}) n_{m} P 
\end{eqnarray}
Finally,  we perform the Jordan-Wigner 
transformation to remove the projector $P$~\footnote{In terms of 
Pauli matrices, the Hamiltonian becomes a
spin model  in an external magnetic field with anisotropic
nearest-neighbor exchange plus the term 
$-\frac{J^2}{8U}(\sigma^{x}_{m+1}\sigma^{x}_{m-1} + \sigma^{y}_{m+1}
\sigma^{y}_{m-1}) 
(\sigma^z_{m} + 1)$.}. Collecting all terms yields the Hamiltonian $H_{F}$
given in Eq.~(\ref{hfm}). This proves the first result announced above.

 Next we consider the continuum limit of $H_{BH}$ and $H_F$.  In both
cases, 
if interactions are neglected, the Hamiltonian is diagonalized by Bloch
waves: $b_{p} = M^{-1/2}_{\rm o} \sum_{m=1}^{M_{\rm o}} e^{-ipx_m} b_{m}$ 
(an identical expression applies to fermions, with $c$'s replacing $b$'s). 
On the lattice, non-interacting particles disperse according to
$\epsilon(p) = -J \cos (pa)$.
Taking $a\to 0$ and expanding $\epsilon(p)$  about $p = 0$
allows  to   identify, by comparing
with the free-particle dispersion $\epsilon_{\rm o}(p) = \hbar^2 p^2/2M$,
the coefficients of the quadratic term: $Ja^2 =  \hbar^2/M$. Therefore, by
keeping the product $Ja^2$ constant as $a\to 0$, one recovers the continuum model
of free particles. In a similar way, after introducing 
the continuum field operators $\Psi(x_m) = a^{-1/2} b_m$, the interaction
term
of $H_{BH}$ becomes $H_{\rm int}= \frac{g}{2} \int_{0}^{L} dx\:
\Psi^{\dag}(x)\Psi^{\dag}(x) 
\Psi(x) \Psi(x)$,   where $g = Ua$ and $L = M_{\rm o} a$  are kept
constant as $a\to 0$. 
Let us now take the $a\to 0$ limit of the interaction term in $H_F$. Using 
$\Psi_{F}(x_m \pm a)  = a^{-1/2} c_{m\pm 1} \approx \Psi_F(x_m) 
\pm a \partial_x \Psi_F(x_m) + a^2 \partial_x^2 \Psi_F(x_m)/2 + O(a^3)$,
where 
$x_m = ma$, and introducing this expression and its hermitian conjugate
into Eq.~(\ref{hfm}),
Eq.~(\ref{hfmc}) is obtained. To close this discussion, it is important to
point out
that the continuum limit which we have taken does not correspond to the
usual continuum 
limit for fermions (see e.g.~\cite{Sachdev00}) where
one linearizes the dispersion around the Fermi points $\pm p_F = \pm \pi
\rho_{\rm o}$. Instead,
we have expanded around the bottom of the band. This is in agreement with
the way in which  the Lieb-Liniger model is obtained from the Bose-Hubbard
Hamiltonian~\footnote{It also
implies neglecting any {\it umklapp} processes, 
which are unphysical for this kind of continuum models.}.

	To obtain the $O(\gamma^{-1})$ correction to the internal
energy of a strongly interacting Bose gas, 
it is convenient to express $H_F$ in Fourier components (henceforth we use
periodic boundary conditions before taking 
the thermodynamic limit $L\to  +\infty$):
\begin{equation}
H_F = \sum_{p} \epsilon(p) c^{\dag}_{p} c_{p} + \frac{g_F}{2L}\sum_{pkq}
(p+q)k \: 
c^{\dag}_{p+q} c^{\dag}_{k-q} c_{k} c_{p}.
\end{equation}
The free energy is obtained from the logarithm of the partition function
($\beta = 1/T$):
\begin{equation}\label{zz}
Z  = {\rm Tr}\: e^{-\beta(H-\mu N)} = Z_{\rm o} \langle{\cal T} \exp\left[
-\frac{1}{\hbar} \int^{\hbar\beta}_{0} d\sigma  \: 
H_{\rm int}(\sigma) \right] \rangle_{\rm o},
\end{equation}
where $H$ is the Hamiltonian of the Lieb-Liniger model and $N$ the total
particle number operator.
In the second expression on the left hand-side $Z_{\rm o} = {\rm Tr} \:
\rho_{\rm o}(\beta,\mu) $,  and
$\langle \cdots \rangle_{\rm o}$ means average with $\rho_{\rm
o}(\beta,\mu) =  e^{-\beta(H_{\rm o}-\mu N)}/Z_{\rm o}$,
i.e. over the non-interacting grand-canonical ensemble. The key to the
calculations that follow is to
realize that  for $\gamma \gg 1$, we can replace  $H$ by $H_F$. Thus,
expanding 
Eq.~(\ref{zz}) in powers of $g_F$  to $O(g_F)$, 
\begin{eqnarray}\label{zz2}
\frac{Z}{Z_{\rm o}} &\simeq& 1 -  \frac{g_F}{2LT} \sum_{pkq} (p+q) k
\langle   
c^{\dag}_{p+q}c^{\dag}_{k-q} c_{k} c_{p} \rangle_{\rm o}   \nonumber \\  
&=& 1 + \frac{g_F L }{2T} \left[\left(\partial_xG(0,\varepsilon)\right)^2
- 
G(0,\varepsilon) \partial^2_xG(0,\varepsilon)\right]. \nonumber \\
\end{eqnarray}
To obtain the last expression we have used Wick's theorem, denoting  the
(time-ordered)
free-fermion  Matsubara-Green's function as $G(x,\sigma)= - \sum_{p}
\langle {\cal T} 
\left[c_{p}(\sigma) c^{\dag}_{p}(0)\right] \rangle_{\rm o}  \: e^{ipx}/L$; 
in the same expression the limit $\epsilon \to 0^{-}$ 
must be taken. The internal energy can be obtained from the above result 
by using the standard expression $U = - \frac{\partial \log
Z}{\partial\beta}  
+ \mu \: \langle N  \rangle$,   which in the present case yields:
\begin{equation}\label{p3}
U =  \left[ U_{\rm o} - \frac{4}{\gamma} \frac{\partial}{\partial \beta} 
\left(\beta U_{\rm o}  \right) + O(\gamma^{-2}) \right],
\end{equation}
where $U_{\rm o} = \sum_{p} \epsilon(p) \langle n_{p}\rangle$ is the 
internal energy of a free Fermi gas (which equals that of a Tonks gas),
and $\langle n_{p} \rangle =
\big( e^{\beta(\epsilon_{\rm o}(p)-\mu')} + 1 \big)^{-1}$ the Fermi-Dirac
distribution.
Using the Sommerfeld expansion~\cite{AM76},  $U_{\rm o}/N_{\rm o}  = 
 \left( \hbar^2 \rho^2_{\rm o}/2M\right)
(\pi^2/3) \left[ 1 + \tau^2/4\pi^2 + O(\tau^4)\right]$, where $\tau =
T/T_{d}$
and $T_{d} = \hbar^2  \rho^2_{\rm o}/2M$ the degeneracy temperature 
in units where $k_{B} = 1$.  Introducing this 
result into Eq.~(\ref{p3}), we obtain for $\tau \ll 1$ and  $\gamma \gg 
1$:
\begin{eqnarray}
\frac{U(T)}{N_{\rm o}} &=& \frac{\hbar^2 \rho_{\rm o}^2}{2M}
\left(\frac{\pi^2}{3} \right)\Big[ 
\left(1 + \frac{\tau^2}{4\pi^2}  \right) \nonumber\\
&& - \frac{4}{\gamma} \left(1 - \frac{\tau^2}{4\pi^2} \right) +  
O(\gamma^{-2}, \tau^4) \Big],
\end{eqnarray}
At zero temperature, this expression reduces to 
$E_{\rm gs} = U(T = 0) =  N_{\rm o}(\hbar^2 \rho^2_{\rm o}/2M) (\pi^2/3)
(1 - 4/\gamma + O(\gamma^{-2}))$,
which agrees to leading order in $\gamma^{-1}$ with the result
obtained by Lieb and Liniger
in Ref.~\onlinecite{Lieb65}. 

	Equation~(\ref{zz2}) also allows us to obtain
asymptotic expressions for the pair-correlation function. As recently
pointed out by Kheruntsyan {\it et al.}~\cite{Kheruntsyan02}, 
for a homogeneous system  this function can be obtained  
using the Hellmann-Feynman theorem:
\begin{equation}
-\frac{1}{\beta} \frac{\partial \log Z}{\partial g} = \frac{L}{2} 
\langle \Psi^{\dag} \Psi^{\dag} \Psi  \Psi \rangle
= \frac{\rho^2_{\rm o} L}{2} g_2.
\end{equation}
Applying this to Eq.~(\ref{zz2}) yields ($\varepsilon \to
0^{-}$):
\begin{equation}\label{g2}
g_{2} = \frac{4}{\gamma^2\rho^{4}_{\rm o}} \left[
\left(\partial_xG(0,\varepsilon)\right)^2 - 
G(0,\varepsilon) \partial^2_xG(0,\varepsilon) \right].
\end{equation}
This provides an explicit derivation of the expression 
employed in Ref.~\onlinecite{Kheruntsyan02}, where Eq.~(\ref{g2})
was used after extrapolating  to finite temperatures 
the $T=0$ result obtained from the Bethe
ansatz wave-function. We believe that the present 
derivation is simpler, and  does not require  to deal 
with the Bethe ansatz solution. Using this expression 
we obtain in terms of $\gamma$ and $\tau = T/T_d$ 
that $g_2(\gamma,\tau) = \frac{4}{3} \left(\pi/\gamma\right)^{2} 
\left[1 + \left( \tau^2/4\pi^2 \right) \right]$, in the quantum degenerate
regime  $\tau \ll 1$, and $g_2(\gamma,\tau) = 2\tau/\gamma^2$
for $1 \ll \tau \ll \gamma^2$, where the gas is non-degenerate but  bosons 
are still strongly interacting.

 Let us finally discuss some of  the phenomena that characterize the 
regime where $K <1$.  We first discuss,
very briefly (further details
will be provided in a future publication~\cite{unpub}), how to enter this regime 
in an optical lattice. In realistic situations,
when $U$ is the largest energy scale of the problem, 
the description provided by Eq.~(\ref{hfm})  is not complete. There is always present
a small repulsion between bosons  in neighboring sites~\cite{Fisher89,Recati02} 
which is described by the operator $V \sum_{m} n_{m }n_{m+1}$, 
with $0 < V \ll U$. When this is taken into account,
the relevant Hamiltonian at large $U$ becomes:
\begin{eqnarray}\label{hfm2}
&&H_F= -\frac{J}{2} \sum^{M_{\rm o}}_{m=1} \left[c^{\dag}_{m+1} c_{m} 
- \frac{J n_m}{U} c^{\dag}_{m+1}c_{m-1}  + {\rm h.c.} \right]
\nonumber \\ 
&& \mbox{       }  + V' \sum_{m=1}^{M_{\rm o}} n_m n_{m+1}  + O\left(\frac{J^3}{U^2}\right),
\end{eqnarray}
where $V' =  V -  J^2/U$. As we argue below, the $K < 1$ regime can be reached 
by increasing $V$. To this purpose, we notice that both $V$ and $U$ are proportional 
to the strength of the interaction between the atoms, which can be tuned to larger values using 
Feshbach resonances. For instance, in an optical lattice near half-filling 
(i.e. for $N_{\rm o} \approx M_{\rm o}/2$) and for sufficiently large $U$, the model
of Eq.~(\ref{hfm2}) can be well approximated~\cite{unpub} by the Heisenberg-Ising chain~\cite{Haldane80}.
For this model, the expression for $K$~\cite{Johnson73} 
obtained from its Bethe-ansatz solution,
\begin{equation}
K = \frac{\pi}{ 2\pi  -2 \cos^{-1}\left( V'/J \right)},
\end{equation}
tells us that $K < 1$ can be achieved by 
making $ 0<V' \lesssim J$. One immediate  consequence of 
the existence of a strong
repulsion with nearest neighbors 
is  the enhancement of density fluctuations. For instance, if the system
is tuned to a Luttinger
liquid  with $K < 1/2$, the density correlation function (which 
behaves as $|x|^{-2K}$ at $T = 0$ and $L \to +\infty$) will
 decay more slowly than the phase correlations 
(behaving as $\sim |x|^{-1/2K}$).  This 
conclusion remains true when the system is confined in a
trap~\cite{Cazalilla02}. 
Furthermore, in a trap the density profile  exhibits Friedel oscillations
of period equal
to  $\rho^{-1}_{\rm o}$  near its boundaries~\cite{Cazalilla02,
Wonnenberg02}. 
The amplitude of these oscillations is  strongly suppressed for $K > 1$, 
and strongly enhanced for $K < 1$. The  higher visibility of the Friedel
oscillations 
can provide an experimental signature 
that this  regime has been reached. By contrast,  in the 
current situation many of the experimentally available quasi-1d systems 
are quasi-condensates ($K > 1$) 
with a fluctuating phase but strongly suppressed density fluctuations.

Furthermore, near half-filling  $N_{\rm o} \approx  M_{\rm o}/2$ and for $V' \gtrsim J$, a 
quantum phase transition to a new type of (boson) Mott insulator  can take
place. The transition is again of Kosterlitz-Thouless type and can be described by a sine-Gordon
model of the form $H = H_{\rm gauss} + g \int dx 
\: \cos\left[4\theta(x) + 4\pi x\delta' \right]$ (see 
e.g.~\cite{Haldane81,Sachdev00}).  
In the insulating ground state ($V' > J$), the lattice occupancy  $n_m$
follows the patterns 
$010101010 \cdots$ or $101010101\cdots$, i.e. the boson density is
modulated 
with twice the lattice period. This could be observed by means of Bragg
scattering. 
The excitations are solitons, interpolating between  the two degenerate
ground states: For 
example $\cdots 010101101010\cdots$ This excitation is fractional and
carries half a boson~\cite{Haldane80,unpub}.
A longitudinal confinement  can be treated as described in
Ref.~\onlinecite{Buchler02}.

  The author acknowledges useful and inspiring conversations with A.
Nersesyan.

\end{document}